\documentclass[aps,12pt,superscriptaddress,amsfonts,amssymb,amsmath]{revtex4}

\usepackage{graphicx}

\begin{document}

\title{
Functional integral transition elements of a massless oscillator} 

\medskip

\author{G.\ Modanese \footnote{Email address: giovanni.modanese@unibz.it }}

\affiliation{Free University of Bolzano, Faculty of Science and Technology, Bolzano, Italy \medskip}

\date{April 22, 2014}

\linespread{0.9}

\begin{abstract}

\medskip

The massless harmonic oscillator is a rare example of a system whose Feynman path integral can be explicitly computed and receives its main contributions from regions of the functional space that are far from the classical and semiclassical configurations near the stationary point of the action. The functional average $\langle q_m^2 \rangle$ of the square of the coordinate at a time $t_m$ which is intermediate between the initial and final time gives a measure of the amplitude of quantum fluctuations with respect to the classical path. This average, or ``transition element'', is divergent in the massless limit, signaling a quantum runaway. We show that the divergence is not due to the continuum limit and formulate the conjecture that the divergent contributions come from regions where the action $S$ is constant and therefore the interference factor $e^{-iS/\hbar}$ does not oscillate. For most systems these regions have zero functional measure and thus give a null contribution to the path integral, but this is not the case for the massless oscillator. We study the simplest functional subspace with constant action, namely the one with $S=0$, which is connected to the classical solutions but extends to infinity, like an hyperplane through the origin; this subspace turns out to be infinite-dimensional. Some possible applications and developments are mentioned.

\end{abstract}


\maketitle

\section{Introduction}

The path integral formulation of quantum mechanics starts from the simple fundamental principle that the total quantum amplitude of any physical process is given by a weighted sum over all possible alternatives. The explicit mathematical implementation of this principle, however, is in general very complex and almost intractable. In spite of considerable efforts \cite{Mazzucchi,Kleinert,PI} our knowledge of Feynman path integrals is far from the level of rigor and insight attained for other physical formalisms. Yet even at the heuristic and purely formal level the path integral technique is very powerful, because it allows clever manipulations of full theories and their perturbative expansions in virtually every field from relativistic high-energy physics to condensed matter physics. Furthermore the Euclidean version, with continuation to imaginary time, leads to rigorous evaluations and approximations thanks to the strong convergence of the real exponential $e^{-S/\hbar}$.

In this work we are concerned with the explicit exact evaluation of path integrals whose main contributions originate from regions of the functional space which are far from the classical and semiclassical configurations near the stationary point of the action. Situations of this kind have been previously studied, in a different context, by da Luz et al.\ \cite{PRA,PHD}. Those works addressed systems with strong constraints or boundary conditions, which lead to a drastic discretization of the configuration space. A very interesting application is to systems, like ``quantum paths'', which are of direct practical interest \cite{q-paths}.

The path integral of a massless oscillator can be regarded as physically  uninteresting, but it is far from trivial and maintains the mathematical complexity of a truly infinite-dimensional functional space. Our aim is to understand the connection between the classical configurations and the dominant ``far'' configurations. In the process, we manage to learn more about certain subtleties of path integrals in general, but also encounter properties which are not completely clear; we formulate some conjectures and elaborate upon them, but a systematic treatment remains elusive.

Our main tool is the explicit evaluation through the path integral of a quantity of the kind called ``transition element'' by Feynman and Hibbs, more precisely the functional average $\langle q_m^2 \rangle$ of the square of the coordinate of the oscillator at a time $t_m$ which is intermediate between the initial and final time. This is not a quantity which is usually evaluated for an oscillator, but it is interesting in our case because it gives a measure of the amplitude of the quantum fluctuations with respect to the classical path, which is simply $q(t)=0$. The divergence of $\langle q_m^2 \rangle$ signals a quantum runaway, in the path integral, from the classical configuration.

The ouline of the paper is the following. In Sect.\ \ref{s1} we recall the case of the familiar oscillator with mass $M$ and compute the transition element $\langle q_m^2 \rangle$ through the known expression for the propagator. We check that in the limit $M \to 0$ the quantum fluctuations diverge. In Sect.\ \ref{s2} we explicitly compute the transition element setting $M = 0$ from the beginning and discretizing the time integral in the action. We re-obtain the divergence and show that it is not due to the continuum limit. We argue that the main contribution to the path integral originates from sectors of the functional space which are far from the classical configuration. We try to understand how this can happen in spite of the interference factor $e^{-iS/\hbar}$, and ``where'' these regions are. 

In Sect.\ \ref{s3} we formulate a conjecture: we suppose that the divergent contributions to the path integral come from regions where $S$ is constant and therefore the factor $e^{-iS/\hbar}$ does not oscillate. For most systems these regions have zero functional measure and thus give a null contribution to the path integral, but we have reasons to believe that this is not the case for the massless oscillator. This conjecture actually emerged in the field-theoretical context of the path integral of the Einstein action; the massless oscillator could be regarded as a toy model for its illustration. 

We study the simplest subspace with constant action, namely the one with $S=0$, which is connected to the classical solution but extends to infinity, like an hyperplane through the origin; this hyperplane indeed turns out to be infinite-dimensional, because in solving the equation $S[q(t)]=0$ we can include in the action an arbitrary function with null integral. Sect.\ \ref{s5} contains our Conclusions and a brief outlook.

\section{Transition element $\langle q_m^2 \rangle$ of the usual harmonic oscillator}
\label{s1}

Let us recall the propagator $K$ of an harmonic oscillator in the case $M \neq 0$, also in order to fix the notation. Let $H$ be the Hamiltonian, given by $H=\frac{1}{2}M{{\dot q}^2} + \frac{1}{2}k{q^2}$, $q_i$ the position at the initial time $t_i$, $q_f$ the position at the final time $t_f$, $T=t_f-t_i$, $S$ the action and $\int d\left[ q \right]$ the functional integral over all possible paths $q(t)$ with the given initial and final conditions. Introducing also the oscillation frequency $\omega=\sqrt{k/M}$, the propagator $K$ is
\begin{equation}
\begin{array}{l}
K({q_f},{t_f};{q_i},{t_i}) = \langle {q_f},{t_f}| {{e^{i\hat{H}({t_f} - {t_i})}}} |{q_i},{t_i}\rangle  = \int {d\left[ q \right]{e^{\frac{i}{\hbar }S\left[ q \right]}}}  = \\
\ \ \ \ \ =\sqrt {\frac{\omega M}{{2\pi i\hbar \sin (\omega T)}}} \exp \left\{ {\frac{{i\omega M}}{{2\hbar \sin (\omega T)}}\left[ {{{\left( {{q_f} + {q_i}} \right)}^2}\cos (\omega T) - 2{q_i}{q_f}} \right]} \right\}
\end{array}
\label{propagator}
\end{equation}
Note that in certain special cases the propagator diverges, for instance when $\omega T=\pi$ , i.e. when $T=\frac{1}{2}T_0$, where $T_0=\frac{2\pi}{\omega}$ is the natural oscillation period of the classical oscillator. This is called the ``caustics'' phenomenon \cite{Kleinert}. The divergence can be interpreted by observing that in that case there exist infinite classical paths connecting the two end-points $q_i$  and $q_f$; these paths have of course different values of the total energy and different velocities at the initial and final points.

In the following we shall consider normalized averages (Feynman-Hibbs ``transition elements'') of the form
\begin{equation}
\left\langle {f(q)} \right\rangle  = \frac{{\int {d\left[ q \right]{e^{\frac{i}{\hbar }S\left[ q \right]}}} f(q)}}{{\int {d\left[ q \right]{e^{\frac{i}{\hbar }S\left[ q \right]}}} }}
\end{equation}
A familiar example in quantum field theory is the correlation functions of the fields:
\begin{equation}
	 	\left\langle {\phi (x)\phi (y)} \right\rangle  = \frac{{\int {d\left[ \phi  \right]{e^{\frac{i}{\hbar }S\left[ \phi  \right]}}} \phi (x)\phi (y)}}{{\int {d\left[ \phi \right]{e^{\frac{i}{\hbar }S\left[ \phi  \right]}}} }}
\end{equation}
where $x$ and $y$ are 4D spacetime coordinates. Here we compute for the oscillator the average of the square of an ``intermediate''  coordinate  $\langle q_m^2 \rangle$. More precisely, consider an intermediate time $t_m$ ($t_i < t_m < t_f$), call $q_m$  the value of the coordinate at the time $t_m$  and compute $\langle q_m^2 \rangle$, with the conditions  $q_i=q_f=0$. Let us break the path integral in two parts, over the intervals  $(t_i,t_m)$ and $(t_m,t_f)$ . We use the propagator formula (\ref{propagator}) to obtain
\begin{equation}
 \begin{array}{l}
\int {d\left[ q \right]{e^{\frac{i}{\hbar }S\left[ q \right]}}} {q_m}^2 = \int {d{q_m}q_m^2K\left( {0,{t_i};{q_m},{t_m}} \right)K\left( {{q_m},{t_m};0,{t_f}} \right)} \\
\int {d\left[ q \right]{e^{\frac{i}{\hbar }S\left[ q \right]}}}  = K(0,{t_i};0,{t_f})
\end{array}
\end{equation}
For simplicity we then suppose that $t_m$  is exactly centered between $t_i$  and $t_f$ , i.e.   ${t_m} - {t_i} = {t_f} - {t_m} = \tau  = T/2$ (but it is easy to generalize the result). We obtain
\begin{equation}
\langle q_m^2 \rangle=
\frac{{\int {d\left[ q \right]{e^{\frac{i}{\hbar }S\left[ q \right]}}} {q_m}^2}}{{\int {d\left[ q \right]{e^{\frac{i}{\hbar }S\left[ q \right]}}} }} = \sqrt {\frac{{M\omega \cos (\omega \tau )}}{{\pi i\hbar \sin (\omega \tau )}}} \int {d{q_m}q_m^2\exp \left\{ {\frac{{i\omega }}{{\hbar \sin (\omega \tau )}}q_m^2\cos (\omega \tau )} \right\}}
\label{formula-qm} 
\end{equation}
It is straightforward to check that the result is 
\begin{equation}
 \langle q_m^2 \rangle=\frac{\hbar}{2i \omega M \cot (\omega \tau)}
=\frac{\hbar}{2i \sqrt{k M} \cot (\omega \tau)}
\label{qm}
\end{equation}
where the factor $\cot (\omega \tau)$ is of order 1, if we suppose that $T$ is of the same magnitude order as the natural oscillation period $T_0$.
We interpret this result as follows. The path integral receives a significant contribution only from those regions in the functional space of the configurations $q(t)$ where the phase factors $e^{iS/\hbar}$ interfere constructively. These are the regions near the stationary point of the action, i.e.\ near the classical solution. With $q_i=q_f=0$ and $T=t_f-t_i \neq \frac{T_0}{2}$, the classical solution is $q(t)=0$. Therefore only small quantum fluctuations near this zero solution are left in the path integral, giving a contribution of order $\hbar$.

This is confirmed by the evaluation of the standard quantum average $\langle 0 \left| \hat{q}^2 \right|0\rangle$ in the ground state of the harmonic oscillator. This average is by construction real and does not correspond to the measurement of an intermediate coordinate like $\langle q_m^2 \rangle$, but still gives a measure of the quantum fluctuations of the $q$ coordinate.
 Using the relation ${E_0} = \frac{1}{2}\hbar \omega  = \frac{1}{2}
\langle 0 \left| \frac{1}{2}M{\omega ^2}{\hat{q}^2} \right|0\rangle$
one finds
 \begin{equation}
\langle 0 \left| \hat{q}^2 \right|0\rangle=\frac{2\hbar}{M\omega}=\frac{2\hbar}{\sqrt{kM}}
\label{0q0}
\end{equation}

From both eq.s (\ref{qm}) and (\ref{0q0}) we can see that when the mass $M$ tends to zero while the spring strength $k$ is kept constant, the amplitude of the fluctuations grows. In the limit of exactly zero mass, we can predict from these equations that $\langle q_m^2 \rangle$ is a divergent quantity. This will indeed be now confirmed by a direct calculation of the path integral in the case $M=0$.

\section{Direct calculation of the transition element for the massless oscillator}
\label{s2}

Let us start directly from the lagrangian of a massless oscillator and compute its path integral. This procedure is quite unusual, because such lagrangian does not have any kinetic term. We are more concerned with mathematics here, however, than with physics. Our aim is to put in evidence a peculiar connection (or rather, disconnection), in this path integral, between the stationary point of the action and the quantum dynamics.

Consider the lagrangian $L=-\frac{1}{2}kq^2$. The classical equation of motion obtained by minimizing the action is just $q(t)=0$. We can also introduce an external force $F(t)$ and rewrite the lagrangian as $L=-\frac{1}{2}kq^2+qF(t)$. The equation of motion becomes $q(t)=F(t)/k$, implying that the oscillator follows the external force, or is at rest at the origin if the external force vanishes. This is the consequence of the complete absence of inertia. The canonical quantization of this system is awkward, because the conjugate momentum $\partial L/\partial \dot q$  is trivially constant; the system is ``constrained'', the energy is not conserved, etc. But let us compute the average $\langle q_m^2 \rangle$ analogous to (\ref{formula-qm}) directly from the path integral, by discretizing the time axis with an infinitesimal parameter $\delta=T/N$. We obtain in the numerator the following ordinary multiple integral over the intermediate coordinates $q_1, \dots ,q_m, \dots ,q_{N-1}$
 \begin{equation}
\int {d\left[ q \right]{e^{\frac{i}{\hbar }S\left[ q \right]}}} {q_m}^2 = \int {\prod\limits_{k = 1}^{N - 1} {d{q_k}} q_m^2\exp \left\{ {\frac{{ik\delta }}{{2\hbar }}\left[ { - q_i^2 - q_1^2... - q_m^2... - q_{N - 1}^2 - q_f^2} \right]} \right\}}
\label{eq10} 
\end{equation}
(Remember that actually $q_i=q_f=0$.) The denominator ${{\int {d\left[ q \right]{e^{\frac{i}{\hbar }S\left[ q \right]}}} }}$ is given by the same integral, without $q_m^2$ in front of the exponential. The exponential is completely factorized and all integrals are simplified in the ratio, except the one in the variable $dq_m$. We obtain
 \begin{equation}
\left\langle {q_m^2} \right\rangle  = \frac{{\int {d{q_m}q_m^2\exp \left\{ {\frac{{ - ik\delta }}{{2\hbar }}q_m^2} \right\}} }}{{\int {d{q_m}\exp \left\{ -{\frac{{ik\delta }}{{2\hbar }}q_m^2} \right\}} }} = \frac{\hbar }{{ik\delta }}
\label{res1}
\end{equation}
In the continuum limit this is divergent. This result is not an anomaly of the continuum limit of our system, because other quantities have finite averages. For comparison, one can evaluate the average of $\exp(-kq_m^2)$, which still gives a Gaussian integral; one easily obtains a finite result, which vanishes in the continuum limit. On the other hand, if we compute the average of a quantity like  $q_m^2/\left( {1 + q_m^2} \right)$, which tends to a finite value for large  $q_m$, we obtain a finite and non-zero result in the continuum limit; to see this, one can exploit the known integral
\begin{equation}
\int {dx\frac{{{x^2}\exp \left( { - \frac{i}{2}a{x^2}} \right)}}{{{x^2} + 1}} = \pi {e^{\frac{{ia}}{2}}}\left\{ { - 1 + {\rm{erf}}\left[ {\left( {\frac{1}{2} + \frac{i}{2}} \right)\sqrt a } \right]} \right\} + \frac{{1 - i}}{{\sqrt a }}\sqrt \pi  } 
\end{equation}
and consider that the ``Error function'' erf($x$) is finite for $a \to 0$, and tends to -1. In our case  $a = const. \cdot \delta $, and $\delta \to 0$  in the continuum limit. The normalization factor at the denominator behaves like $\frac{1}{{\sqrt \delta  }}$, therefore for $\delta \to 0$  the term $\frac{{1 - i}}{{\sqrt a }}\sqrt \pi  $  gives a finite contribution and the rest goes to zero. In conclusion, the quantum averages computed with the path integral tell us consistently that each intermediate coordinate $q_m$  tends to grow without limit. Quantum-mechanically, the system runs away to large $q$, far from the classical solution. Actually, the divergence of $\left\langle {q_m^2} \right\rangle$ for $M=0$ had been already predicted from the ``explosion'' of the fluctuations in eq.s (\ref{qm}), (\ref{0q0}).

\section{The ``hyperplane'' $S=0$ in the space of configurations}
\label{s3}

The crucial question now is: which configurations in the path integral contribute to this divergent result? Certainly not those near the stationary point of the action, $q(t)=0$. We make the following conjecture: {\it The main contributions to the path integral come from regions of the functional configuration space in which $S$ is constant, although not stationary, in such a way that these contributions interfere constructively.}

For most usual dynamical systems these regions with constant action, reminiscent of equipotential surfaces in a force field, have lower dimensionality and zero measure, so they cannot really contribute to the path integral. We shall show, however, that in the case of the massless oscillator they are full-dimensional subspaces of the configuration space.

As a first step towards the general case, we give a characterization of these regions in the zero action case $S=0$. The region with zero action in functional space is somewhat easier to visualize, because it is ``connected'' to the classical trajectory $q(t)=0$, i.e.\ geometrically to the origin of space. We can picture this region as an infinite-dimensional hyperplane through the origin; for all the configurations $q(t)$ lying on this hyperplane the action is constantly zero, but it changes if we move out of the hyperplane, since only the origin is a stationary point where $\delta S=0$ in any direction. The value of $\langle q^2_m \rangle$ is not limited on the hyperplane.

The functional integration on this hyperplane yields a significant contribution to the path integral, because the functional measure of this set is not zero and there is no destructive interference between neighboring paths. This contribution is weighed with the overall constant factor $e^{ iS_0/\hbar}=1$ ($S_0=0$), both in the numerator and denominator of (\ref{formula-qm}). Note that if this was the only significant contribution to $\langle q^2_m \rangle$ from the total path integral, then $\langle q^2_m \rangle$ would be real, while the explicit calculation shows that it is imaginary. This means, if our conjecture has to be consistent, that in the total path integral there are other non-classical contributions of the same kind, proportional to overall factors $e^{iS_1/\hbar}$, $e^{iS_2/\hbar} \dots $ ($S_1, S_2 ... \neq 0$), and which do not simplify in the ratio $\int d[q]e^{iS/\hbar} {q_m}^2/\int d[q]e^{iS/\hbar}$. It is possible to write some algebraic relations which must hold among these contributions; this will be done in a forthcoming paper.

\section{Proof that the hyperplane $S=0$ is infinite-dimensional}
\label{s4}

In order to show that the hyperplane $S=0$ is infinite-dimensional we first need  to relax the conditions $q_i=q_f=0$ on the paths. This may look counterintuitive for the classical paths, but is technically allowed in the path integral for the transition element $\langle q^2_m \rangle$ because the result does not change if we add in the numerator and denominator an integration over $q_i$ and $q_f$. Being the exponential completely factorized, we can just rewrite $\int dq_i \int dq_f \int {d\left[ q \right]{e^{\frac{i}{\hbar }S\left[ q \right]}}} {q_m}^2$ instead of $ \int {d\left[ q \right]{e^{\frac{i}{\hbar }S\left[ q \right]}}} {q_m}^2$ in eq.\ (\ref{eq10}), and similarly for the denominator $ \int {d\left[ q \right]{e^{\frac{i}{\hbar }S\left[ q \right]}}} $.

Next we look for configurations $q(t)$ such that $L(q)=0$, without the constraint $q_i=q_f=0$. For this we exploit another known invariance property of the path integral: its value does not depend on the presence in the lagrangian of an additive total time derivative of a function of $q$ and $\dot{q}$. If, for instance, we add to the originary lagrangian $L =  - \frac{1}{2}k{q^2}$ a term $\mu \dot{q}$, with $\mu$ constant, it is clear that the classical dynamics remains the same and one can check explicitly that also the result for $\left\langle {q_m^2} \right\rangle$ in (\ref{res1}) does not change. Therefore we can write
\begin{equation}
\mu \dot q - \frac{1}{2}k{q^2} = 0
\label{eq-non-def}
\end{equation}
and look for solutions of this differential equation with arbitrary boundary conditions at the times $t_i$ and $t_f$. We obtain
\begin{equation}
q(t) =  - \frac{1}{{\frac{{kt}}{{2\mu }} + c}}
\label{solution}
\end{equation}
with $c$ arbitrary constant. In order to fix the ideas, take ${\frac{{k}}{{2\mu }}}=1$ and suppose that the time interval is $(t_i,t_f)=(0,1)$. By giving the constant $c$ a small arbitrary value $c=\varepsilon >0$, we obtain in the interval $(0,1)$ finite solutions which grow from the value $-1/\varepsilon$ in $t=0$ to the value $ - \frac{1}{{ - 1 + \varepsilon }} \cong  - 1$ in $t=1$. 

\begin{figure}
\begin{center}
  \includegraphics[width=10cm,height=7cm]{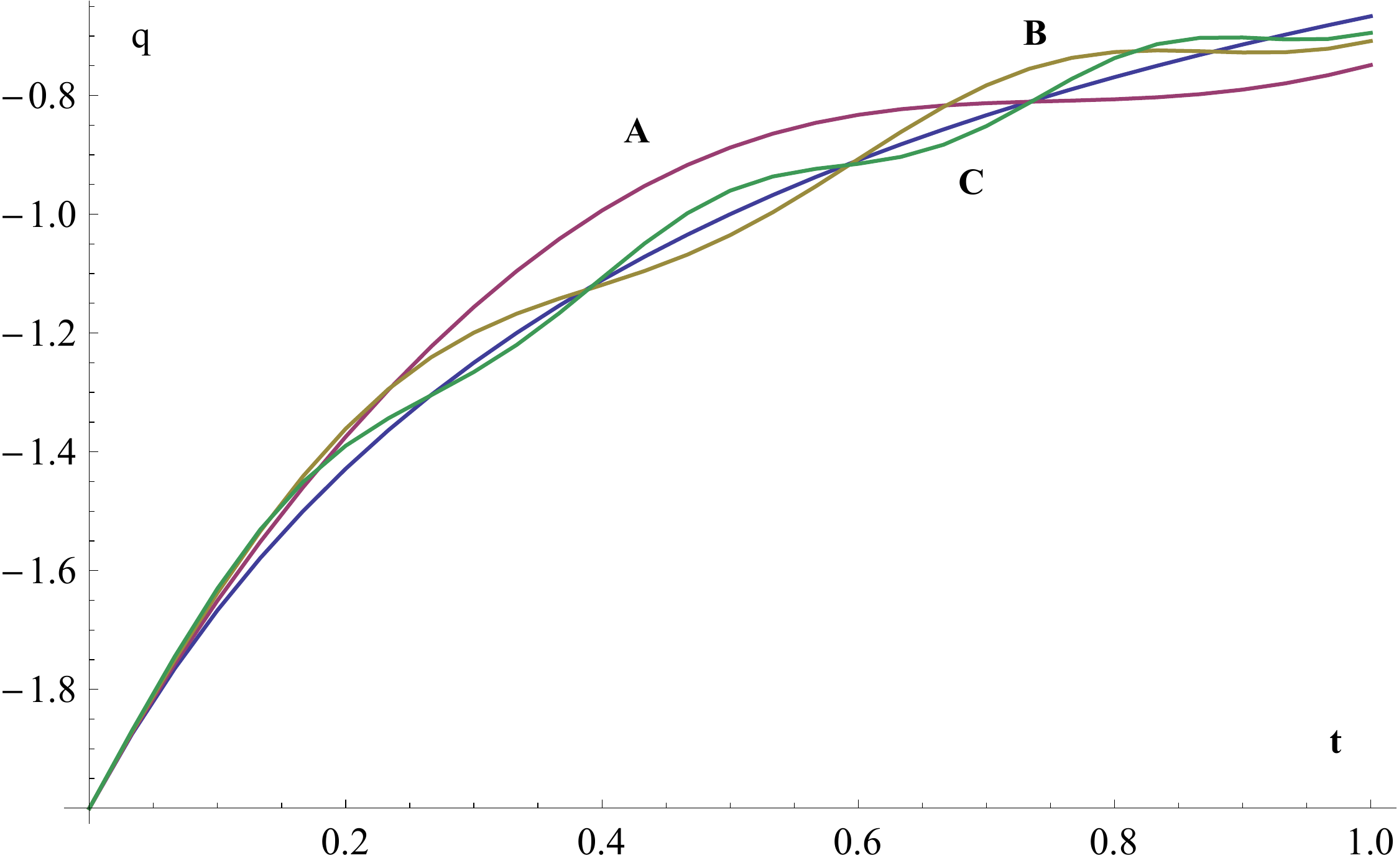}
\caption{Solutions of eq.s (\ref{eq-def}), (\ref{gamma}) with $n=0$ (undeformed) and with $n=1,2,3$ (lines A, B, C). The initial condition is $q(0)=-2$.} 
\end{center}      
\end{figure}

These solutions display non-bounded values of $q$ and depend on the parameter $\varepsilon$, so they certainly are elements of the hyperplane $S=0$, but do not constitute yet an infinite-dimensional space. They satisfy, however, the ``punctual'' condition $L(q)=0$, while it is actually sufficient to satisfy the integral condition $S = \int\limits_0^1 {dtL(q) = 0} $. Therefore we look for solutions, instead of (\ref{eq-non-def}), of the ``deformed'' equation (with ${\frac{{k}}{{2\mu }}}=1$)
\begin{equation}
\dot q - {q^2} = g(t)
\label{eq-def}
\end{equation}
where $g(t)$ is a generic continuous function with null integral in $(0,1)$. We can take, for instance,
\begin{equation}
g(t)= \gamma \sin (2\pi n t), \ \ \ \ n \in {\bf{N}}
\label{gamma}
\end{equation}
where $\gamma$ is thought of as a finite but small perturbative parameter. General theorems about ordinary differential equations ensure the existence of a finite solution in $(0,1)$ for a certain range of values of $\gamma$. It is immediate to generate numerically some of these solutions, for instance with $\gamma=0.6$ and $n=1,2,3...$ (Fig.\ 1). These functions are only slightly deformed with respect to the solution (\ref{solution}), but constitute now really an infinite-dimensional space, corresponding to all the possible variations of the function $g(t)$. In this space, $S[q]$ is zero and $\left\langle {q_m^2} \right\rangle$ is not bounded. This completes the proof of the existence.

\section{Conclusions and outlook}
\label{s5}

The non-interference mechanism which generates the main contributions to the present functional integral can be generalized to Quantum Field Theory. The massless oscillator can be regarded as a model for its illustration. For instance, the existence of ``zero-modes'' of the Einstein action has been proven in \cite{BUC}. An explicit calculation of their contribution to the path integral is quite complex and will be presented in a forthcoming paper.

Note that although the path integral does not depend on the insertion of the $\dot q$-term in the action, the ``positioning'' in the functional space of the subspaces with constant action depends on it. For example, it is clear that if we set the minimal action $S =  - \frac{1}{2} \int dt \, k{q^2}$ equal to zero, we cannot obtain any non-trivial solutions $q(t)$. It follows that in that case the regions contributing to the path integral must have $S \neq 0$, do not contain the origin and are disconnected from the classical configuration $q(t)=0$.

The general invariance of the functional integration with respect to changes in the total-derivative terms in the lagrangian also appears to deserve further analysis.


\begin{thebibliography}{100}

\bibitem{Mazzucchi}
Mazzucchi, S., {\it Mathematical Feynman path integrals and their applications}, World Scientific, 2009

\bibitem{Kleinert}
Kleinert, H., {\it Path Integrals in Quantum Mechanics, Statistics, Polymer Physics and Financial Markets}, World Scientific, 2009

\bibitem{PI}
Casalbuoni, R., et al., Ed.s, {\it Path Integrals from peV to TeV}, World Scientific, 1999

\bibitem{BUC}
Modanese, G., The vacuum state of quantum gravity contains large virtual masses, Class. Quant. Grav. {\bf 24} (2007) 1899

\bibitem{PRA}
Da Luz, M.G.E., and Bin Kang Cheng, Quantum-mechanical results for a free particle inside a box with general boundary conditions, Phys. Rev. A {\bf 51} (1995) 1811-1819

\bibitem{PHD}
Ozorio de Almeida, A.M., and M.G.E. da Luz, Path integrals and edge corrections for torus maps, Physica D {\bf 94} (1996) 1-18

\bibitem{q-paths}
Viswanathan, G.M., et al., Optimizing the success of random searches, Nature {\bf 401} (1999) 911-914





\end{thebibliography}
\end{document}